\documentclass[11pt]{article}
\usepackage{graphicx}

\title{A constraint on antigravity
of antimatter from precision spectroscopy of simple atoms}

\author{Savely G. Karshenboim\\
~\\
Max-Planck-Institut f\"ur Quantenoptik, Garching, 85748, Germany\\
D.~I. Mendeleev Institute for Metrology, St.Petersburg, 190005,
Russia\\
{\em email\/}: savely.karshenboim@mpq.mpg.de}

\begin{document}

\maketitle

\begin{abstract}
Consideration of antigravity for antiparticles is an attractive
target for various experimental projects. There are a number of
theoretical arguments against it but it is not quite clear what kind
of experimental data and theoretical suggestions are involved. In
this paper we present straightforward arguments against a
possibility of antigravity based on a few simple theoretical
suggestions and some experimental data. The data are: astrophysical
data on rotation of the Solar System in respect to the center of our
galaxy and precision spectroscopy data on hydrogen and positronium.
The theoretical suggestions for the case of absence of the
gravitational field are: equality of electron and positron mass and
equality of proton and positron charge. We also assume that QED is
correct at the level of accuracy where it is clearly confirmed
experimentally.
\end{abstract}

\section*{Introduction}

After producing cold antihydrogen atoms via the recombination of
positrons and antiprotons \cite{atrap,athena} a number of
possibilities to experimentally test CPT invariance including
precision spectroscopy and gravitation have been intensively
discussed. In particular, a question of experimental check for a
possibility for antigravity of antiparticles rose for new
consideration.

It is absolutely clear that the very suggestion of antigravity for
antimatter contradicts general relativity (GR). While physics of
matter objects in free fall cannot be distinguished from the zero
gravity case (as long as any gradient effects can be neglected), an
antigravitating particle will clearly recognize the gravitation
field.

Here we refer to antigravity for the following situation:
\begin{itemize}
\item[(i)] the inertial masses of particles and antiparticles are the
same;
\item[(ii)] the gravitation masses, understood, e.g., as the
related coefficients for Newtonian gravity at rest for long
distances (where the Newtonian gravity strongly dominates over any
general relativity effects), have the same absolute values for
particles and antiparticles;
\item[(iii)] the particle--particle and
antiparticle--antiparticle gravitational interaction is attractive,
while the particle--antiparticle one is repulsive.
\end{itemize}
In particular, that means that in any given [weak] gravitation
field a particle and its antiparticle experience forces, equal in
their absolute values and opposite in directions. A concise review
on the issue of antigravity can be found, e.g., in \cite{nieto}.

General relativity sets an even stronger direct constraint on a
possibility of antigravity. Observation of various effects beyond
Newtonian gravity \cite{turyshev} proves that gravitation is due to
a tensor field which (in contrast to a vector field for
electromagnetic forces) produces only attraction and no repulsion.
Any speculation on the so-called fifth force considers only small
corrections beyond the tensor forces \cite{equivalence} and can be
ignored in the case of a 100\% effect such as antigravity.

Still, because of the sensitivity of this issue, the experimental
community would prefer to have some more straightforward constraint,
for which it would be clear which experimental data and theoretical
suggestions are involved in ruling out antigravity.

Here we present such a consideration.

Prior to a detailed examination, we have to make two remarks. First,
we remind that there are two kinds of gravitation-related effects.
Some are sensitive to a certain long-range difference of the
gravitation potential $U({\bf r})$, while the other are sensitive to
the local field strength ${\bf g}=-{\bf \nabla}U({\bf r})$.

Second, when one `tests' gravity, a few kinds of interactions are
commonly considered. Some, like true gravitation, couple to the mass
$m$, while the others couple to some charges. The charge is additive
and it is the same, e.g., for the ground and excited states of an
atom. We do not consider here interactions coupled to such
quantities as magnetic moments, which are neither additive, nor
conserved. The mass and charges are regular. Mass is a
state-dependent property in atomic physics since different states
possess different binding energy. This energy is directly related to
emission/absorption photon energy. On the contrary, any conserved
charge is state-independent and it is the same for any state in the
same atom. The magnetic moment (or a similar quantity) is
state-dependent, but it depends on the internal structure of a
related state.

In particular, antigravity means a coupling to mass. Otherwise, we
have to suggest that all the gravity is a force coupled to a
baryonic and/or leptonic charge. The latter indeed changes its sign
for antimatter, but the results for tests of the equivalence
principle and GR in general indeed rule out this option
\cite{turyshev,equivalence}.

\section*{The red shift of the photon frequency as the blue shift of
the clocks}

The idea of our derivation is based on a study of the red shift in
atoms and we remind here how one can derive the red shift expression
with `minimal' and clearly formulated suggestions. (Note, once we
consider a possibility for antigravity of antiparticles we cannot
blindly rely on GR.)

Let us consider ground and excited states of an atom in two
positions at different gravitational potentials. Different states
have different masses such as
\begin{equation}
\Delta m c^2 = h\nu_0 \;,
\end{equation}
where $\nu_0$ is the photon frequency. The recoil effects for
emission or absorption are neglected and the related correction does
not change the result of our consideration.

\begin{figure}[ptbh]
\begin{center}
\includegraphics[height=2.5cm,clip]{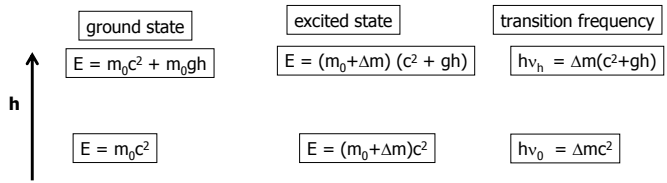}
\end{center}
\caption{Derivation of the gravitational red shift.\label{f:red} }
\end{figure}

With masses being different for the ground and excited states, the
gravitation energy is also different (see Fig.~\ref{f:sens})
\begin{equation}
\Delta E = \Delta m \;\Delta U
\end{equation}
by a value which is related to a correction for the
emitted/absorbed photon frequency
\begin{equation}
h\Delta \nu = \Delta m \Delta U \;.
\end{equation}
Within this consideration the shift mentioned is for the frequency
of a clock, while a photon traveling through the time-independent
gravitation field does not change its frequency\footnote{Actually,
it is rather natural to suggest (once we assume antigravity of
antimatter) that the matter and antimatter are affected by gravity
in opposite directions, while `pure neutral objects' such as
positronium atoms and photons are not affected by gravity at all.
That immediately leads to a contradiction with gravitational
deflection of light, which is well established experimentally.}. The
derivation above, which follows \cite{okun}, shows that the
gravitational red shift is not a specific property of GR, but rather
a generic property of any relativistic theory of gravity since it is
supposed to reproduce Newtonian gravity and special relativity as
crucial limits.

What is called the gravitational red shift of the photon frequency
within GR depends on a choice of 4-coordinates. Depending on how the
time is defined (or how clocks are synchronized in space), it may be
a shift in photon frequency or in clock frequency (or in both). The
observable effect is a certain mismatch in photon frequency while
communicating between two clocks at different gravitational
potentials. In particular, in our consideration, instead of the red
shift of the photon going `upward' we observe a blue shift of the
transition frequency in the upper clock in respect to the bottom
clock (see, e.g., \cite{okun}). The mismatch of the frequency sent
to an upper potential is equivalent to the red shift by a
conventional value:
\begin{equation}
\frac{\Delta \nu}{\nu_0} = \frac{\Delta U}{c^2} \;.
\end{equation}

Usually the red-shift experiments are organized as relative and
differential. We can either compare the red shift of two different
clocks, or we can compare the experimental red shift to a
theoretical calculation, based on a measurement of the free fall
acceleration with bulk matter or atomic beams. In the former case we
check universality of the red shift for different traveling
frequencies and different frequency sources. Actually, this is the
universality of the gravitational effect on all local clocks that
allows to redefine the time in any way we like. The latter case
confirms that the gravitation acting on matter governs photon
behavior as well. In both cases the gravitational red shift is a
comparison of a certain frequency of a clock measured at two
different positions.

Looking for a non-universality of the red shift in different clocks
we have a certain advantage. Usually, one also has to deal with a
number of motional effects, which partly or completely cancel
gravitational effects under certain conditions. Studying a
differential red shift (i.e., a difference in red shifts of two
clocks), the motional effects, which are always universal,  do not
contribute.

\section*{Potential of a gravitational attractor for an orbital frame}

Let us check another option. Instead of measuring at two location we
can measure a ratio of two frequencies at one location at a certain
distance $r$ from a point-like attractor and determine the same at
an infinite distance from gravitating masses theoretically. In this
case, we can take advantage of considering a weak field, which may
accumulate a large value of $\Delta U$ over a large distance, while
the field by itself is negligible for any ${\bf g}$-related effects.

A circular rotation of a probe body with velocity $v$ around mass
$M$ deals with the field (acceleration)
\begin{equation}
 g=\frac{{ v}^2}{r}=G\frac{M}{r^2}
\end{equation}
and the difference of potential (in respect to an infinitely
remote point) is equal to
\begin{equation}
\Delta U(r) = -G\frac{M}{r}=-{v}^2\;.
\end{equation}
That in particular means that if we need to estimate any
$U(r)$-related effect, we can do a proper estimation once we know
the velocity of rotation. The non-circularity of the orbit cannot
change the order of magnitude for the estimation.

Let us consider a scenario with a slowly rotating body at a circular
orbit. Effects due to the local gravitational force are small as
long as the distance is large. However, the potential in respect to
an infinitely remote zero-gravity point is not small. If we position
our probe clocks on this body, we will observe a gravitational shift
with respect to infinitely remote points as
\begin{equation}
\frac{\Delta \nu}{\nu_0} = -\frac{v^2}{c^2} \;.
\end{equation}

In our laboratory experiments we deal with a few motions which
produce relatively large values of ${v^2}/{c^2}$. For our motion
around the Sun we have ${v^2}/{c^2}\sim 10^{-8}$, while the Solar
System moving around the center of our galaxy supplies us with a
value ${v^2}/{c^2}\sim 10^{-6}$ \cite{galaxy}. We do not expect that
a non-circularity of the orbits can change the estimations. We note
that the related field is very small, but the potential difference
is large enough. Indeed, the universe is not so simple as consisting
of a gravitating center which determines solar and Earth motion
mentioned. However, a good approximation is that most of the
universe is homogenously distributed around us and its gravitational
effect vanishes. (In any case, any other motion induced by a
gravitational potential at a larger scale can be treated in the same
way.)

If we consider only matter with the equivalence principle for
gravitational and inertial masses ($m_g=m_i$), the shift is
universal and cannot be locally detected since all clocks are
shifted in the same way.

\section*{Comparison of hydrogen and antihydrogen and positronium clocks}

Once we suggest antigravity for antimatter and include antimatter
clocks into consideration, the matter and antimatter clocks would be
shifted in opposite directions. Once we know the ratio of two
frequencies at $r=\infty$, we can compare it with a local value and
prove or disprove antigravity.

There is no antimatter clock available for the moment, but a kind of
a `neutral' clock is available. Once we suggest antigravity for
antimatter, the positronium is a system with no gravity at all.

In particular, assuming the antigravity once should expect
\begin{eqnarray}
\nu (r) = \nu_0 \times
 \left\{
 \begin{array}{cc}
 \left(1+\frac{U(r)-U(\infty)}{c^2}\right)              &\mbox{for H}\\
 1   &\mbox{for Ps} \\
  \left(1-\frac{U(r)-U(\infty)}{c^2}\right)              &\mbox{for $\overline{H}$}
 \end{array}
 \right.\;.
\end{eqnarray}
Meanwhile, the frequencies $\nu_0$, unperturbed by gravity, are the
same for the same transition in the hydrogen and antihydrogen atoms,
while the ratio for positronium and hygrogen is calculable (see
below).

We note that the positronium $1s-2s$ transition
\cite{pos1s2sa,pos1s2sb} was measured with a high accuracy as well
as the $1s-2s$ \cite{h1s2s} and some other \cite{h_other}
transitions in hydrogen. And that can be applied to experimental
searches for the antigravity option.

The ratio of the $1s-2s$ frequencies in hydrogen and positronium can
be calculated at zero gravity. In the leading approximation
(Schr\"odinger equation for a particle in the Coulomb field) this
ratio is equal to the ratio of the related reduced masses which are
\begin{eqnarray}
m_{\rm R}({\rm Ps})&=&\frac{m_e}{2}\;,\nonumber\\
m_{\rm R}({\rm H})&=&\frac{m_e}{1+m_e/m_p}\;,\nonumber
\end{eqnarray}
where $m_e$ is the electron mass and $m_p$ is the proton mass. A
number of corrections are known (see, e.g., \cite{review}) and the
theoretical result for the unperturbed frequency ratio has the form
\begin{equation}
\frac{\nu_0^{\rm Ps}(1^3S_1-2^3S_1)}{\nu_0^{\rm
H}(1^3S_1-2^3S_1)}=\frac{1}{2}\,\frac{1}{1+m_e/m_p}\times
F\bigl(\alpha,m_e/m_p\bigr)\;,
\end{equation}
where $F\bigl(\alpha,m_e/m_p\bigr)$ stands for known QED corrections
(see \cite{review,review1}.

The theoretical ratio agrees with the experimental ratio within an
uncertainty of a few parts in $10^9$.

\begin{figure}[hbtp]
\begin{center}
\includegraphics[width=.7\textwidth]{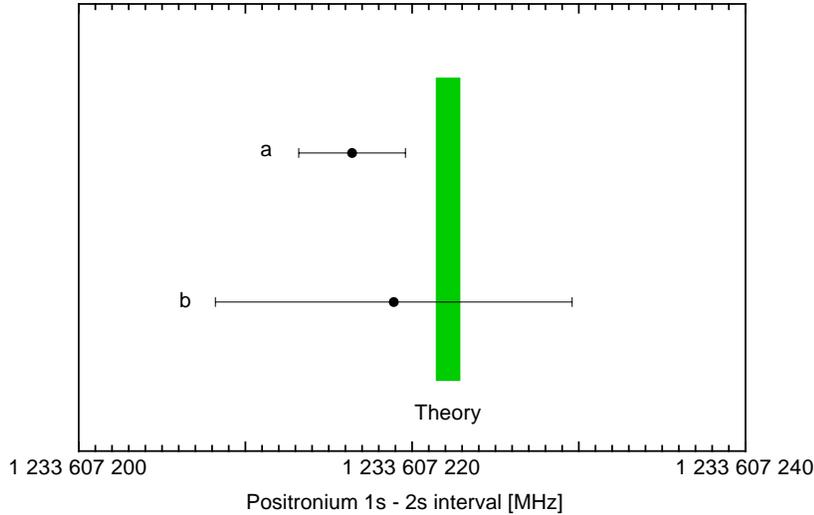}
\end{center}
\caption{\label{F1s} Determination of the $1^3S_1-2^3S_1$ interval
in positronium. The references here are: $a$ is from \cite{pos1s2sa}
and $b$ is from \cite{pos1s2sb}. The plot is taken from
\protect\cite{review}.}
\end{figure}

Technically, instead of considering the frequency ratio for
hydrogen and positronium transitions, we compare positronium
experimental results for $\nu^{\rm th}(1^3S_1-2^3S_1)$ with theory
based on the value of the Rydberg constant obtained from the
hydrogen spectroscopy (see, e.g. \cite{mohr}). A comparison of
theory and experiment is plotted in Fig.~1. The theoretical result
(see, e.g., \cite{review})
\begin{equation}
\nu_0^{\rm th}(1^3S_1-2^3S_1)=1\,233\,607\,222.2(6)\;\mbox{Hz}
\end{equation}
perfectly agrees with experiment \cite{pos1s2sa,pos1s2sb}
\begin{eqnarray}
\nu^{\rm a}(1^3S_1-2^3S_1)&=&1\,233\,607\,219(11)\;\mbox{Hz}\;,\nonumber\\
\nu^{\rm
b}(1^3S_1-2^3S_1)&=&1\,233\,607\,216.4(32)\;\mbox{Hz}\;.\nonumber
\end{eqnarray}

In the absence of gravity, theory should agree with experiment (once
the calculations are correct). However, in the presence of gravity
and under the suggestion of antigravity of antimatter we should
introduce the gravitational corrections. That will affect the
hydrogen theory and shift the value of the Rydberg constant from
hydrogen spectroscopy by $\Delta U/c^2$.

Such a correction should obviously shift a theoretical value and
produce a discrepancy between theory and experiment at the level of
one ppm. That immediately rules out any room for antigravity once we
accept that the galactic field which determines rotation of the
Solar system does not include any exotic component. That is at the
level of a few hundred standard deviations.

Here we considered a rotation around the center of galaxy as only
motion. If we consider some more intergalaxy-scale motions
\cite{galaxy}, that will add an additional gravitational source and
will increase the effect. However, that will not change the order of
magnitude of the effect.

If we address only the solar gravity and use the orbiting velocity
of Earth, we arrive at a limitation at the level of $4\sigma$ which
opposes a possibility for antigravity, but is not too convincing.

\section*{QED and antigravity}

The theoretical ratio is obtained once we assume that the Coulomb
interaction of an electron and a positron is the same as that of a
proton and an electron. The other assumption applied is that the
inertial masses of an electron and a positron are the same.

We can rely on the fact that there are many direct or indirect
tests for these two facts at the level of interest ($10^{-6}$).
However, we prefer just to mention that obviously any `exact'
compensation between the gravitational red shift and inequalities
of masses and charges could happen only if they are of
gravitational origin, which can be checked in many other ways.

In any case an agreement for $g-2$ experiments with uncertainty at
the level of a few parts in $10^{11}$ confirms that an old-fashioned
Dirac equation, which suggests the identity of mass and charge
values for electron and positron, is still a reasonably good
approximation. The results for $g-2$ are of two kinds: there are (i)
measurements on slow electrons and positrons which confirm that both
agree with each other \cite{slow} and with theory and (ii) an
experiment on ultrarelativistic beams of electrons and positrons
\cite{fast}, which produces the same result for either beams.

To be more specific with comparison of the most accurate electron
$g-2$ experiment \cite{g2exp} with theory \cite{g2th}, we note
that for a successful application of theory one needs a value of
the fine structure constant $\alpha$. Instead of comparison of
$g-2$ theory and experiment one can equivalently compare different
alpha's. The most accurate $\alpha$ is from electron $g-2$
\cite{g2exp,g2th}, the second in accuracy method is for atomic
interferometry with cold rubidium \cite{alpha-at-rb} and caesium
\cite{alpha-at-cs} beams. These results agree (see \cite{mohr} for
detail) at the level of uncertainty much lower than $10^{-6}$.

These arguments mean that we can really rely on equality of electron
and positron masses and on universality of elementary charge as the
value of charge of electron, positron, proton etc.

As it was mentioned, from the theoretical viewpoint antigravity is
rather unlikely. Can the method considered here be used for any
other constraints? It seems that not in any straightforward way.
E.g., if the $1s-2s$ transition in antihydrogen were measured with
about the same accuracy as in hydrogen, which means an uncertainty
at the level of a part in $10^{14}$, combining that with a $10^{-6}$
red shift we could claim universality of free fall for matter and
antimatter at the uncertainty level of $10^{-8}$. However, that
would not be very useful. That would be correct only for
`mass-related' gravity. Only the latter interferes with the red
shift since the excited and ground states have different mass. Once
we suggest an exotic coupling due to, e.g., baryonic charge, that
would affect `mechanical' free fall (i.e., a real fall of a bulk
body or an atom), but not its small contribution to the red shift,
since the baryonic charge in ground and excited states is the same.

Nevertheless, in general, considering a particular model of
modification of gravity, some constraint on its parameters could be
set and the transitions of interest are transitions measured in
hydrogen, positronium, muonium and antiprotonic helium (see below)
and transitions in antihydrogen which are to be measured in some
future.

In principle, there may be some reservations about the application
of the galaxy gravitation field that is produced by matter and a
substantial amount of the dark matter. Still, consideration of the
gravitation of the Sun at the Earth orbit is also sufficient to rule
out antigravity. In any case, the uncertainty due to the nature of
the Sun gravity cannot be larger than for the Earth gravity where
the antihydrogen antigravity experiments are supposed to be
realized.

As already mentioned, we could also consider muonium and
antiprotonic helium. Muonium consists of a heavy nucleus, which is a
positive muon (an antiparticle), and a light orbiting particle,
which is an electron. Antiprotonic helium consists of a conventional
nucleus (the $\alpha$ particle) and an orbiting antiproton and an
electron. The latter can be treated as a perturbation. The related
results are quite accurate \cite{muonium,ahelium}, but it is hard to
interpret them in a model-independent way for an antigravity search.
In a conventional case we deal only with inertial mass of a compound
system, which is a sum of component masses corrected due to their
binding energy. The latter is related to the whole two-body system
and well defined. Shares of a binding energy related to each
component are not observable. If we assume antigravity for
antiparticles, we should split the binding energy between the
material and antimaterial components of the system. Such a share can
be indeed assigned {\em ad hoc\/}, but since it is not observable in
any other effect, there is no way to prove any particular
assignment. However, we note that the pair of atoms mentioned,
namely muonium and antiprotonic helium, have a complementary
structure: both consist of two components (once we neglect an
electron in antiprotonic helium) with a particle and an antiparticle
present; one component is light, the other is heavy and both options
are realized: heavy particle and heavy antiparticle.

In principle, the electromagnetically bound two-body system is
just the simplest example of an electromagnetic problem for a
particle or antiparticle bound by the electromagnetic field
produced by other particles or antiparticles.

In particular, we can consider resonance transitions between
quantized levels of particles and antiparticles in the magnetic
field, which we indirectly mentioned when we referred to $g-2$
experiments as a kind of clock. Indeed, the field is produced by
matter and it is a difficult question how to split a `material' and
`antimaterial' part of the binding energy of an antiparticle in the
magnetic field. Considering the transitions in terms of a search for
the gravitational red shift, which is potentially much higher than
the uncertainty in some experiments, we can find a controversy
considering electrodynamics with electrons and positrons in the free
falling frame. Further consideration needs indeed a model.

However, if we like to consider electrodynamics, we should
introduce the electromagnetic field and thus the energy levels
mentioned above, e.g., the Landau levels are to be determined by
the field value. Because of the gravitational red shift, the
result of the field action on a particle in the case of
antigravity could potentially depend on the origin of the field
(whether it is created by matter or antimatter) and kind of a
probe particle (whether it is a particle or antiparticle) at the
ppm level in laboratory experiments. Antigravity would make
classical electrodynamics quite problematic.

In the case of an antiparticle at a classically produced
electromagnetic field we arrive at the same problem -- how to split
the binding energy between the material field source  and a particle
of antimatter, which is crucial for antigravity. It seems that the
most consistent way would be to distribute it according to the
masses of the objects. E.g., for equal charges the share of binding
energy between two inertial masses $m_i$ could be proportional to
$m_i/(m_1+m_2)$. In such a case muonium would behave in the
gravitational field nearly as an atom of antimatter (since $m_\mu\gg
m_e$), while in antiprotonic helium-4, the gravitational mass would
be about 60\% of the inertial mass ($0.6=(4-1)/(4+1)$). For the
antiprotonic helium-3, also studied experimentally \cite{ahelium}
and theoretically \cite{korobov}, the gravitational mass would be
50\% of the inertial mass.

We remind that the experimental uncertainty of the $1s-2s$ in
muonium \cite{muonium} is 4 ppb, while the combined uncertainty of
antiprotonic helium spectroscopy \cite{ahelium} is about 60 ppm
and it is going to be substantially improved. Theory has
sufficient accuracy (see, e.g., \cite{review,review1,korobov}).
Applying both we should also rule out the antigravity option
considering rotation around the center of galaxy.

\section*{Conclusion}

Concluding the paper, we summarize that we have presented
straightforward arguments against a possibility of antigravity
based on a few simple theoretical assumptions and experimental
data. The data are: astrophysical data on rotation of the Solar
System in respect to the center of our galaxy and precision
spectroscopy data on hydrogen and positronium. The theoretical
assumptions for the case of absence of gravitational field are:
equality of electron and positron mass and equality of proton and
positron charge. We also assume that QED is correct at the level
of accuracy where it is clearly confirmed experimentally.

For future activity, we have to mention one more option for a
possible model-independent limitation on antigravity. If a
measurement of the $1s-2s$ transition in positronium can be
improved by an order of magnitude (or a measurement of this
transition in hydrogen will be performed at the same accuracy),
one should be able to observe such a clear signature of
antigravity as annual variation of the ratio of the $1s-2s$
transitions in hydrogen and positronium (or antihydrogen) due to
change in distance between the Sun and the Earth which changes by
about 5 million kilometers during the year. That is related to a
change in the gravitational potential in fractional units by
\begin{equation}
\frac{\Delta U(r_{\rm max})-\Delta U(r_{\rm min})}{c^2}\simeq
3.2\times10^{-10}\;.
\end{equation}
The shift of all material clocks (including hydrogen's) is a blue
shift for $r=r_{\rm max}$ in respect to $r=r_{\rm min}$, while a
positronium clock would experience no shift and the antihydrogen
clocks would be red shifted.

\begin{figure}[ptbh]
\begin{center}
\includegraphics[height=6cm,clip]{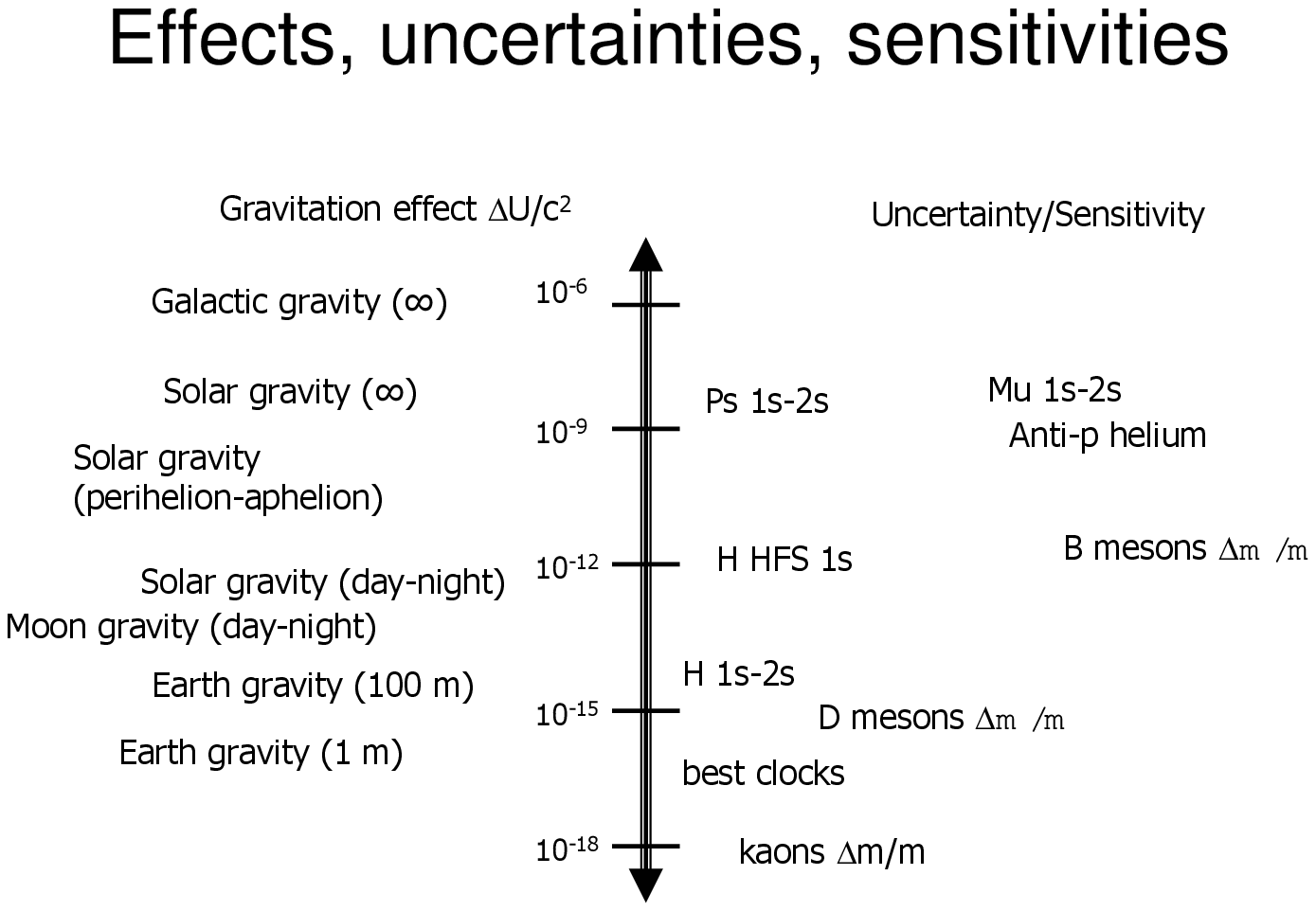}
\end{center}
\caption{Fractional values of gravitational effects versus
uncertainty and sensitivity of various precision measurements (see
also \cite{kaon}).\label{f:sens} }
\end{figure}

There is a number of experiments with low uncertainty or with a high
sensitivity and a summary is presented in Fig.~\ref{f:sens}.

Due to importance of positronium spectrum and annihilation of free
and bound positrons study of positronium annihilation line from the
Galactic Centre region can also deliver some constraints on
antigravity. At present the accuracy (as observed by SPI/INTEGRAL
\cite{integral}) is not sufficient but may be improved in future
missions.

\section*{Acknowledgments}

The work was in part supported by DFG (under grant \# GZ 436 RUS
113/769/0-3) and RFBR (under grants \# 08-02-91969, 06-02-16156,
08-02-13516).

Stimulating discussions with M. Fujiwara, J. Walz, D. A.
Varshalovich, R. A. Sunyaev, S. I. Eidelman, and V. G. Ivanov are
gratefully acknowledged.

\end{document}